\documentclass[aps,twocolumn,showpacs]{revtex4}
\usepackage[brazil]{babel}
\usepackage{graphicx}
\usepackage{color}
\usepackage[latin1]{inputenc}
\usepackage[T1]{fontenc}
\usepackage[brazil]{babel}

\RequirePackage{amsfonts,amssymb,amsmath}

\begin{document}
\title{Quantum Singularities Around a Global Monopole}
\author{Jo\~ao Paulo M. Pitelli} 
\email{e-mail:pitelli@ime.unicamp.br}
\author{Patricio S. Letelier} 
 \email{e-mail: letelier@ime.unicamp.br} 
\affiliation{
Departamento de Matem\'atica Aplicada-IMECC,
Universidade Estadual de Campinas,
13081-970 Campinas,  Sao Paulo, Brazil}

\begin{abstract}
The behavior of a massive scalar particle on the spacetime surrounding a monopole is studied from a quantum mechanical point of view. All the boundary conditions necessary to turn the spatial portion of the wave operator self-adjoint are found and their importance to the quantum interpretation of singularities is emphasized.

\end{abstract}

\maketitle

\section{Introduction}
\label{introduction}
Classical singularities in general relativity are indicated by incomplete geodesics or incomplete paths of bounded acceleration \cite{ellis}. They are classified into three basic types : A singular point $p$ in the spacetime is a quasiregular singularity if no observer sees any physical quantities diverging even if its world line reaches the singularity. A singular point $p$ is called a scalar curvature singularity if every observer that approaches the singularity sees physical quantities such as tidal forces and energy density diverging. Finally, in a nonscalar curvature singularity, there are some curves in which the observers experience unbounded tidal forces \cite{helliwell2,konkowski}. Because the spacetime is by definition differentiable, points representing singularities must be excluded  from our manifold. The geodesic incompleteness leads to the lack of predictability of the future of a classical test particle which reaches the singularity.

It is this lack of predictability that links classical and quantum singularity. Analogous to the classical case, we say that a spacetime is quantum mechanically singular if the evolution of a wave function representing a one particle state is not uniquely determined by the initial state. That is to say that we need a boundary condition near the singularity in order to obtain the time evolution of the wave packet~\cite{horowitz}. An example of a classical  singularity  which becomes nonsingular with the introduction of quantum mechanics is the hydrogen atom. Solving the Schr\"odinger equation for the Coulomb potential  equation and imposing  square-integrability of the solutions is enough to obtain a complete set of solutions which span $L^2(\mathbb{R}^3)$.  Then  the evolution of the initial wave packet is uniquely determined. There are others examples of classically singular spacetimes that become nonsingular in  view of quantum mechanics~\cite{horowitz,helliwell3}. But, unfortunately, there are much more examples of spacetimes which remain singular~\cite{helliwell2,ishibashi, pitelli1,pitelli2}.

In this paper we will study the spacetime of a global monopole from a quantum mechanical point of view. 
We believe that  this  will be the first time that the ideas of Horowitz and Marolf~\cite{horowitz} are applied to a non vacuum solution of Einstein equations, since the metric of a global monopole due to Barriola and Vilenkin~ \cite{barriola} (deficit of a solid angle)  represents  a solution of the Einstein field  equations  with spherical symmetry with  matter that extends to infinity (cloud of cosmic strings with spherical symmetry~\cite{letelier}). This point will be discussed in the next section.

This paper is organized as follows: In Sec. \ref{metricglobal} we briefly review the spacetime of a global monopole. In Sec. \ref{quantum} we explore the definition of quantum singularities and present the criterion that will decide whether the spacetime is quantum mechanically singular or not. In Sec. \ref{resultados} we use the methods described in  Sec. \ref{quantum} to the case of the global monopole. Finally in Sec. \ref{discussion} we discuss the implications of the results of Sec. \ref{resultados}.

\section{The Metric of a Global Monopole}
\label{metricglobal}
One of the predictions of the grand unification theories is the arising of topological defects. They are produced during the phase transitions in the early universe and their existence is a very attractive scenario for large scale structure formation; see, for instance, \cite{vilenkin,vilenkin2}. The most simple example of a topological defect is the cosmic string; see, for instance, \cite{helliwell}. This defect  appears in the breaking of a $U(1)$ symmetry group and has the metric,
\begin{equation}
ds^2=-dt^2+d\rho^2+\alpha^2\rho^2d\varphi^2+dz^2.
\label{metricstring}
\end{equation}
The factor $\alpha^2$ in the metric (\ref{metricstring}) introduces a deficit angle $\Delta=2\pi(1-\alpha)$ on a spatial section $z=\text{const}$. 
This metric is characterized by a null Riemann-Christoffel curvature tensor everywhere except on a line ($z$ axis), where it is proportional to a Dirac delta function; see \cite{letelier2}. The energy-momentum tensor associated with (\ref{metricstring}) is 
\begin{equation}
T_{t}^{t}=T_{z}^{z}=\frac{(1-\alpha)}{4\alpha\rho}\delta(\rho),\label{stringemt}
\end{equation}
i. e., for strings we have  the equation of state:  energy density  equal to tension. 

Barriola and Vilenkin \cite{barriola} considered a monopole as associated
 with a triplet of scalar fields $\phi^{a}$ ($a=1,2,3$) given by
\begin{equation}
\phi^a=\eta f(r)x^a/r,
\end{equation}
with $x^ax^a=r^2$ and $\eta$ is the spontaneous symmetry breaking scale. 

Considering the most general static metric with spherical symmetry,
\begin{equation}
ds^2=-B(r)dt^2+A(r)dr^2+r^2(d\theta^2+\sin^2{\theta}d\phi^2),
\end{equation}
and that outside the monopole the function $f$ takes the value $1$,
they obtain  the  solution of the Einstein equations,
\begin{equation}
ds^2=-dt^2+dr^2+\alpha^2r^2(d\theta^2+\sin^2{\theta}d\phi^2),
\label{metric}
\end{equation}
where $\alpha^2=(1-8\pi G\eta^2)$.

Note that on the hypersurface given by $\theta=\frac{\pi}{2}$, the spacetime is conical \cite{helliwell} and geodesics on this surface behaves as geodesics on a cone with angular deficit $\Delta=2\pi(1-\alpha)$. This fact can mislead us to  think  that this is  an empty flat spacetime with a topological defect. But this is not the case, as we will see in the following paragraphs.

In a previous   work \cite{letelier}, one of the authors found the metric (\ref{metric}) following a completely different approach, the search of  the metric associated with a cloud of cosmic strings with 
spherical symmetry. This spacetime is not isometric to Minkowski spacetime (as is the spacetime surrounding a cosmic string \cite{helliwell}) since we have a   non zero   tetradical  component
 of the curvature tensor,
\begin{equation}
R_{\hat{\theta}\hat{\phi}\hat{\theta}\hat{\phi}}=\frac{1-\alpha^2}{\alpha^2r^2},
\label{riemann}
\end{equation}
nonvanishing everywhere, opposed to the curvature in the spacetime of a cosmic string, which is proportional to a Dirac delta function with support on the string \cite{letelier2}. 
{The spacetime around a monopole has a scalar curvature singularity, since all observers who approach the singularity will see physical quantities, such as tidal forces, diverging \cite{helliwell2}.
Note that the Ricci scalar is twice the value of $R_{\hat{\theta}\hat{\phi}\hat{\theta}\hat{\phi}}$.

Despite the fact that the  curvature tends to zero when $r\to\infty$, the spacetime of a global monopole is not asymptotically flat. The energy-momentum tensor $T_{\mu\nu}$ has only the components $T_{t}^{t}=T_{r}^{r}=\eta^2/r^2$ that are  nonvanishing everywhere. A similar situation due to the existence of closed timelike curves was discussed  in  \cite{misner}. In the present case  it arises due to the slow falling off of $T^{t}_{t}$ that  has as a consequence  the divergence of $M(r)=\int_{0}^{r}{T^{t}_{t}(r)r^2\text{d}r}$ as $r\to\infty$.

The Newtonian potential $\Phi=GM(r)/r$ defined  in Ref. \cite{barriola} is constant and since the $T_{t}^{t}$ component of the energy-momentum tensor is given by $T_{t}^{t}=\eta^2/r^2$, we have $M(r)$ proportional to the distance $r$. But, in view of the Poisson equation, such a constant potential is inconsistent with a nonvanishing density, as noted by Raychoudhuri \cite{raychoudhuri}. Since  $T_{t}^{t}\propto 1/r^2$ we have that  the correct expression for the potential is $\Phi\sim \ln{r}$ with gravitational intensity $\propto 1/r$. So there is in fact a Newtonian gravitational force on the matter around the monopole~\cite{raychoudhuri}.

It is clear that we can not interpret the  metric (\ref{metric}) as representing an isolated massive object introduced into a previously flat  universe. So it must be regarded as a symmetric cloud of cosmic string, with all the string forming the cloud  intersect in a single point $r=0$.
\section{Quantum Singularities}
\label{quantum}
Classical singularities can be interpreted via quantum mechanics by using the definition of Horowitz and Marolf, who considered a classically singular spacetime as quantum mechanically nonsingular when the evolution of a general state is uniquely determined for all time \cite{horowitz}; in other words, that the spatial portion of the wave operator is self-adjoint.

The wave operator for a massive scalar field is given by 
\begin{equation}
\frac{\partial^2 \Psi}{\partial t^2}=-A\Psi,
\label{scalarfield}
\end{equation}
where $A=-VD^i(VD_i)+V^2M^2$ and $V=-\xi_{\mu}\xi^{\mu}$, with $\xi^{\mu}$ being a timelike Killing vector field and $D_i$ the spatial covariant derivative on a static slice $\Sigma$. Let us choose the domain of $A$ to be $\mathcal{D}(A)=C_{0}^{\infty}(\Sigma)$ in order to avoid the singular points. In this way $A$ is a symmetric positive definite operator, but this domain is so small, so restrictive that the adjoint of $A$, i.e., $A^{\ast}$ is defined on a much larger domain $\mathcal{D}(A^{\ast})=\{\psi\in L^2:A\psi\in L^2\}$ and $A$ is not self-adjoint. In order to transform the operator $A$ into self-adjoint one we must extend its domain until the domains $\mathcal{D}(A)$ and $\mathcal{D}(A^{\ast})$ are equal. If the extended operator $(\overline{A},\mathcal{D}(\overline{A}))$ is unique~\footnote{In this case the closure of A. See Refs. \cite{reed} and \cite{richtmeyer}. } then $A$ is said to be essentially  self-adjoint and the evolution of a quantum test particle obeying (\ref{scalarfield}) is given by
\begin{equation}
\psi(t)=exp(-it\overline{A}^{1/2})\psi(0).
\end{equation}
The spacetime is said to be quantum mechanically nonsingular. Otherwise there is one specific evolution for each self-adjoint extension $A_{E}$
\begin{equation}
\psi_{E}(t)=exp(-itA_{E}^{1/2})\psi(0)
\end{equation}
and the spacetime is quantum mechanically singular.
The criterion used to determine if an operator  is essentially self-adjoint comes from a theorem by von Neumann \cite{reed}, which says that the self-adjoint extensions of an operator $A$ are in one-to-one correspondence with the isometries from $Ker(A^{\ast}-i)$ to $Ker(A^{\ast}+i)$. We solve the equations
\begin{equation}
A^{\ast}\psi\mp i\psi=0 
\label{maisoumenos}
\end{equation}
and count the number of linear independent solutions in $L^2$. If there is no solution for the above equations, then $\text{dim}(Ker(A^{\ast})\mp i)=0$, and if there are no isometries from $Ker(A^{\ast}-i)$ to $Ker(A^{\ast}+i)$, the operator is essentially self-adjoint. Otherwise, if there is one solution for each equation (\ref{maisoumenos}), there is a one-parameter family of isometries and therefore a one-parameter family of self-adjoint extensions and so on.
\section{Quantum singularities on the global monopole background}
\label{resultados}

From the metric (\ref{metric}) and the identity 
\begin{equation}
\square \Psi=g^{-1/2}\partial_{\mu}\left[g^{1/2}g^{\mu\nu}\partial_{\nu}\right]\Psi
\end{equation}
we have that the Klein-Gordon equation reads
\begin{equation}\begin{aligned}
\frac{\partial^2\Psi}{\partial t^2}=&\frac{1}{r^2}\frac{\partial}{\partial r}\left(r^2\frac{\partial \Psi}{\partial r} \right)+\frac{1}{\alpha^2r^2\sin{\theta}}\frac{\partial}{\partial \theta}\left(\sin{\theta}\frac{\partial \Psi}{\partial \theta}\right)\\&+\frac{1}{\alpha^2r^2\sin^{2}{\theta}}\frac{\partial^2\Psi}{\partial \varphi^2}-M^2\Psi.
\label{principal}
\end{aligned}
\end{equation}

From  (\ref{scalarfield}) we find 
\begin{equation}\begin{aligned}
-A=&\frac{1}{r^2}\frac{\partial}{\partial r}\left(r^2\frac{\partial}{\partial r} \right)+\frac{1}{\alpha^2r^2\sin{\theta}}\frac{\partial}{\partial \theta}\left(\sin{\theta}\frac{\partial }{\partial \theta}\right)\\&+\frac{1}{\alpha^2r^2\sin^{2}{\theta}}\frac{\partial^2}{\partial \varphi^2}-M^2\Psi
\end{aligned}
\end{equation}
and the equation to be solved is
\begin{equation}\begin{aligned}
(A^{\ast}\mp i)\psi=&\frac{1}{r^2}\frac{\partial}{\partial r}\left(r^2\frac{\partial \psi}{\partial r} \right)+\frac{1}{\alpha^2r^2\sin{\theta}}\frac{\partial}{\partial  \theta}\left(\sin{\theta}\frac{\partial \psi }{\partial \theta}\right)\\&+\frac{1}{\alpha^2r^2\sin^{2}{\theta}}\frac{\partial^2 \psi }{\partial \varphi^2}+(\pm i-M^2) \psi.
\end{aligned}
\end{equation}
 We shall  closely follow Sec. III of Ref. \cite{ishibashi}, where an illustrative example of a flat spacetime with a point removed (texture) is studied.

By separating  variables, $\psi=R(r)Y_{l}^{m}(\theta,\varphi)$, we get the radial equation
\begin{equation}
\frac{d^2R(r)}{dr^2}+\frac{2}{r}\frac{dR(r)}{dr}+\left[(\pm i-M^2)-\frac{l(l+1)}{\alpha^2r^2}\right]R(r).
\label{equacao}
\end{equation}

Let us first consider the case $r=\infty$. The above equation takes the form
\begin{equation}
\frac{d^2R(r)}{dr^2}+\frac{2}{r}\frac{dR(r)}{dr}+(\pm i -M^2)R(r)=0,
\end{equation}
whose solution is
\begin{equation}   
R(r)=\frac{1}{r}[C_1e^{\beta r}+C_2e^{-\beta r}],
\label{solucaoinfinito}
\end{equation}
where 
\begin{equation}
\beta=\frac{1}{\sqrt{2}}\bigg[(\sqrt{1+M^4}+M^2)^{1/2} \mp 
i (\sqrt{1+M^4}-M^2)^{1/2}\bigg].
\end{equation}

Obviously, solution (\ref{solucaoinfinito}) is square-integrable near infinity if and only if $C_{1}=0$. Then the  asymptotic behavior of $R(r)$ is given by $R(r)\sim\frac{1}{r}e^{-\beta r}$.

Near $r=0$, Eq. (\ref{equacao}) reduces to 
\begin{equation}
\frac{d^2R(r)}{dr^2}+\frac{2}{r}\frac{dR(r)}{dr}-\frac{l(l+1)}{\alpha^2r^2}R(r)=0,
\end{equation}
whose solution is $R(r)\sim r^{\gamma}$, where 
\begin{equation}
\gamma=\frac{-1\pm\sqrt{1+4\frac{l(l+1)}{\alpha^2}}}{2}.
\end{equation}

For $\gamma=-\frac{1}{2}+\frac{1}{2}\sqrt{1+4\frac{l(l+1)}{\alpha^2}}$ the solution $R(r)\sim r^{\gamma}$ is square-integrable near $r=0$, that is, 
\begin{equation}
\int_{0}^{\text{constant}}{|r^{\gamma}|^2r^2dr}<\infty. 
\end{equation}
And for $\gamma=-\frac{1}{2}-\frac{1}{2}\sqrt{1+\frac{l(l+1)}{\alpha^2}}$ we have
\begin{equation}
\int_{0}^{\text{contant}}{|r^\gamma|^2r^2dr}=\int_{0}^{\text{contant}}{r^{1-\sqrt{1+4\frac{l(l+1)}{\alpha^2}}}} dr.
\end{equation}
Therefore in order for $r^{\gamma}$ be square integrable we have that
\begin{equation}
1-\sqrt{1+4\frac{l(l+1)}{\alpha^2}}>-1
\end{equation}
and
\begin{equation}
l(l+1)<\frac{3}{4}\alpha^2.
\end{equation}
This condition is satisfied only if $l=0$. In fact, the mode $R_0(r)$ does not belong to $L^2(\mathbb{R}^3)$ because $\nabla^2(1/r)=4\pi\delta^3(r)$. But, we have a physical singularity at $r=0$ so  $r=0 \notin \Sigma$ and $R_{0}(r)$ is an allowed mode.  Then near origin we have
\begin{equation}
R_0(r)=\tilde{C}_1+\tilde{C}_{2}r^{-1}
\label{near0}
\end{equation} 
and we can adjust the constants in Eq. (\ref{near0}) to meet the asymptotical behavior $R_{0}(r)\sim e^{-\beta r}$~\footnote{We say that $r=\infty$ is in the point-limit case and $r=0$ is in the circle-limit case. When this happens, we can assure that there is one solution of Eq. (\ref{maisoumenos}). See Refs. \cite{reed} and \cite{richtmeyer}.}. There is one solution for each sign in Eq. (\ref{maisoumenos}), so there is a one-parameter family of self-adjoint extensions of $A$.

In order to better understand this result, let us solve exactly equation (\ref{principal}) using a separation of variables of the form
\begin{equation}
\Psi(t,r,\theta,\varphi)=e^{-i\omega t}R(r)Y_{l}^{m}(\theta,\varphi).
\end{equation}
For the radial equation we have
\begin{equation} 
r^2R''(r)+2rR'(r)+[(k^2r^2-l(l+1)/\alpha^2]R(r)=0,
\label{equacaor}
\end{equation}
where $k^2=\omega^2-M^2$.
By doing $u\equiv kr$
we find
\begin{equation}
u^2R''(u)+2uR'(u)+[u^2-l(l+1)/\alpha^2]R(u)=0.
\end{equation}
And defining  $Z(u)$ by
\begin{equation}
R(u)=\frac{Z(u)}{u^{1/2}},
\end{equation}
we get
\begin{equation}
Z''(u)+\frac{1}{u}Z'(u)+\left\{1-\frac{1}{4u^2}\left[1+4\frac{l(l+1)}{\alpha^2}\right]\right\}Z(u)=0
\end{equation}
which is the Bessel equation of order $\delta_{l}$,
\begin{equation}
\delta_{l}=\frac{1}{2}\sqrt{1+4\frac{l(l+1)}{\alpha^2}}.
\end{equation}

Therefore the  general solution of Eq. (\ref{equacaor}) is
\begin{equation}
R_{\omega,l,m}(r)=A\frac{J_{\delta l}(kr)}{\sqrt{kr}}+B\frac{N_{\delta l}(kr)}{\sqrt{kr}}.
\label{solution}
\end{equation}
Let us analyze the square-integrability near $r=0$ of each one of the functions appearing in  the above equation.

The Bessel functions are always square-integrable near the origin, i.e.,
\begin{equation}
\int_{0}^{\text{constant}}{\left|\frac{J_{\delta_{l}}(kr)}{\sqrt{kr}}\right|^{2}r^{2}dr}<\infty.
\end{equation}
The behavior of  the  Newmann functions near the origin is given by $N_{\nu}(x)\propto \left(\frac{2}{x}\right)^{\nu}$. Therefore,
\begin{equation}
\int_{0}^{\text{constant}}{\left|\frac{N_{\delta_{l}}(kr)}{\sqrt{kr}}\right|^{2}r^2dr}\sim\int_{0}^{\text{constant}}{r^{-2\delta_{l}+1}} 
\end{equation}
so that 
\begin{equation}
-2\delta_{l}+1>-1\Rightarrow\delta_{l}<1\Rightarrow\sqrt{1+4\frac{l(l+1)}{\alpha^2}}<2 \Rightarrow l=0
\end{equation}
in order to be square integrable.

Therefore, for $l\neq 0$ square integrability suffices to determine uniquely solution (\ref{solution}), while for $l=0$ we need an extra boundary condition. To find this extra boundary condition, let us define $G(r)=rR_{0}(r)$. Then Eq. (\ref{equacaor}) with $l=0$ becomes
\begin{equation}
\frac{d^2G(r)}{dr^2}+k^2G(r)=0.
\end{equation}

The boundary condition for the above equation is simple   (see Refs. \cite{ishibashi} and \cite{richtmeyer}) and it is given by
\begin{equation}
G(0)-aG'(0)=0.
\end{equation}

Therefore $G(r)$ we have
\begin{equation}
G(r)\propto\left\{\begin{aligned} &\cos{kr}+\frac{1}{ak}\sin{kr}\;\;\; &a\neq 0\\
&\cos{kr}&a=0\end{aligned}\right.
\label{condcont}
\end{equation}

The general solution of Eq. (\ref{principal}) is 
\begin{equation}
\Psi_{a}=\int{\text{d}\omega e^{-i\omega t}\sum_{l=0}^{\infty}\sum_{m=-l}^{l}C(\omega,l,m)R_{\omega,l,m}(r)Y_{l}^{m}(\theta,\varphi)},
\end{equation}
with
\begin{equation}
R_{\omega,l,m}=\frac{J_{\delta_{l}}(kr)}{\sqrt{kr}}\;\;\; l\neq0
\end{equation}
and
\begin{equation}
R_{\omega,0,0}(r)=
\left\{\begin{aligned}
&\frac{\cos{kr}}{r}+\frac{1}{ak}\frac{\sin{kr}}{r}\;\;\;&a\neq0\\
&\frac{\cos{kr}}{r} & a=0
\end{aligned}\right.
\label{parameter a}
\end{equation}

\section{Discussion}
\label{discussion}

In the previous section we found [Eq. (\ref{parameter a})] a one-parameter family of solutions of Eq. (\ref{principal}), each one corresponding to a determined value of $a\in\mathbb{R}$.  The theory does not tell us how to pick up one determined solution, or even if there exists such a distinguished one. Any solution is as good as the others, so the spacetime of a global monopole (or of a cloud of strings with spherical symmetry to be more precise) remains singular in the view of quantum mechanics. The future of a given initial wave packet obeying the Klein-Gordon equation is uncertain, as well as it is uncertain of the future of a classical particle which reaches the classical singularity in $r=0$.

\acknowledgements 
We thank Fapesp for financial  support and  P.S.L also thanks  CNPq.


\end{document}